\documentclass[amsmath,amssymb,showkeys,superscriptaddress,aps,prd,10pt,onecolumn]{revtex4-2}
\usepackage{graphicx}
\usepackage[utf8]{inputenc}
\usepackage[caption=false]{subfig}
\usepackage{multirow}
\usepackage{appendix}
\usepackage{graphicx,epstopdf}

\usepackage{colortbl}
\usepackage{xcolor}
\usepackage[colorlinks=true, urlcolor=blue, linkcolor=blue, citecolor=blue]{hyperref}
\usepackage{float}

\begin{document}

\title{Unveiling the Structure of  Hidden-Bottom Strange Pentaquarks via Magnetic Moments}
\author{Halil Mutuk}%
\email[]{hmutuk@omu.edu.tr}
\affiliation{Department of Physics, Faculty of Sciences, Ondokuz Mayis University, 55200 Samsun, Türkiye}

\author{Xian-Wei Kang}%
\email[]{xwkang@bnu.edu.cn}
\affiliation{Key Laboratory of Beam Technology of the Ministry of Education, College of Nuclear Science and Technology, Beijing Normal University, Beijing 100875, China}
\affiliation{Institute of Radiation Technology, Beijing Academy of Science and Technology, Beijing 100875, China}

 
\begin{abstract}
Motivated by the discovery of hidden-charm strange pentaquarks, we conduct a systematic study of the magnetic moments of the hidden-bottom strange  pentaquarks in molecular picture. We calculate magnetic moments of hidden-bottom strange pentaquarks with strangeness-1 and 2. Magnetic moment gives valuable information about the inner structure and shape of the hadron. The obtained results may be helpful to determine the inner structure of these  hypothetical states.
\end{abstract}

\maketitle

\section{Motivation}\label{introduction}

Although there were theoretical predictions about the existence of multiquark states apart from conventional hadrons $(q\bar{q}, qqq)$, the discovery of $\chi_{c1}(3872)$ by the Belle Collaboration in 2003 \cite{Belle:2003nnu} opened a new era in the hadron physics. This particle was the first observed four-quark state.  A large number of non-conventional hadrons have been observed by the experimental facilities in the past two decades. These non-conventional hadronic states are called exotic hadrons, since their physical features cannot be fit to the quark model.

The paradigm of multiquark states turned into a more interesting path when the pentaquark states are observed. In 2015, the LHCb Collaboration announced the observation of two peaks in the $J/\psi p$ invariant mass spectrum of $\Lambda_b^0 \to J/\psi p K^-$ process with masses and widths \cite{LHCb:2015yax}:
\begin{eqnarray}
m_{P_c(4380)^+}&=&4380 \pm8 \pm 29~\mathrm{MeV}, \quad \Gamma_{P_c(4380)^+}=205 \pm 18 \pm 86~\mathrm{MeV}, \\
m_{P_c(4450)^+}&=&4449.8 \pm 1.7 \pm 2.5~\mathrm{MeV}, \quad \Gamma_{P_c(4450)^+}= 39 \pm 5 \pm 19~\mathrm{MeV}.
\end{eqnarray}

Using an updated data, in 2019 LHCb Collaboration reported three new pentaquark states named as $P_c(4312)$, $P_c(4440)$ and $P_c(4457)$ \cite{LHCb:2019kea}. The previously reported $P_c(4450)$ state was resolved into two narrow states as $P_c(4440)$ and $P_c(4457)$. The masses and widhts of these new pentaquark states are  as follows
\begin{eqnarray}
m_{P_c(4312)^+}&=&4311.9 \pm 0.7^{ +6.8}_{-0.6}~\mathrm{MeV}, \quad \Gamma_{P_c(4312)^+}=9.8 \pm 2.7 ^{ +3.7}_{-4.5}~\mathrm{MeV},\\
m_{P_c(4440)^+}&=&4440.3 \pm 1.3 ^{+ 4.1}_{-4.7}~\mathrm{MeV}, \quad \Gamma_{P_c(4440)^+}= 20.6 \pm 4.9^{+8.7}_{-10.1}~\mathrm{MeV},\\
m_{P_c(4457)^+}&=&4457.3 \pm 0.6 ^{+ 4.1}_{-1.7}~\mathrm{MeV}, \quad \Gamma_{P_c(4457)^+}= 6.4 \pm 2.0^{+5.7}_{-1.9}~\mathrm{MeV}.
\end{eqnarray}
According to exotic hadron naming convention \cite{Gershon:2022xnn}, these three states are denoted as 
\begin{equation}
P_c(4312) \to P_{\psi}^N(4312), ~ P_c(4440) \to P_{\psi}^N(4440), ~ P_c(4457) \to P_{\psi}^N(4457).
\end{equation}

In 2021, LHCb reported evidence of the $P_{cs}(4459)$ ($P_{\psi s}^{\Lambda}(4459)$ in the new naming convention) with charm quark content by analyzing of $J/\psi \Lambda$ invariant mass distribution in $\Xi_b^-\rightarrow J/\psi K^-\Lambda$ decays~\cite{LHCb:2020jpq} with the mass and width
\begin{equation}
m_{P_{\psi s}^{\Lambda}(4459)}=4458.8 \pm 2.9^{+4.7}_{-1.1}~\mathrm{MeV}, \quad \Gamma_{P_{\psi s}^{\Lambda}(4459)} = 17.3 \pm 6.5^{+8.0}_{-5.7}~\mathrm{MeV},
\end{equation}
respectively. Very recently, $P_{\psi s}^{\Lambda}(4338)$ $(P_{cs}(4338))$ state is observed in the $B^{-}\rightarrow J/\psi \Lambda \bar{p}$ process with mass and width
\begin{equation}
m_{P_{\psi s}^{\Lambda}(4338)}=4338.2 \pm 0.7\pm 0.4~\mathrm{MeV}, \quad \Gamma_{P_{\psi s}^{\Lambda}(4338)} = 7.0\pm 1.2 \pm 1.3~\mathrm{MeV},
\end{equation}
respectively, \cite{LHCb:2022ogu}.
The spin-parities of $P_{\psi s}^{\Lambda}(4459)$ and $P_{\psi s}^{\Lambda}(4338)$ states are preferably $\frac{3}{2}^{-}$ and $\frac{1}{2}^{-}$.

After the observation of hidden-charm strange pentaquark states, various studies with different models are devoted to understand the spectroscopic parameters such as by using effective field theories \cite{Du:2021bgb, Zhu:2021lhd, Feijoo:2022rxf, Zhu:2022wpi, Yan:2022wuz, Chen:2022wkh,Chen:2022onm,Wang:2023ael, Wang:2023iox,Nakamura:2022gtu,Yalikun:2023waw,Wu:2024lud}, line shapes \cite{Meng:2022wgl,Nakamura:2022gtu}, QCD sum rules~\cite{Wang:2022gfb, Wang:2020eep, Chen:2020uif, Wang:2022neq, Azizi:2023foj}, phenomenological quark models~\cite{Shi:2021wyt, Xiao:2021rgp, Ortega:2022uyu, Giachino:2022pws, Yang:2022ezl, Wang:2022mxy, Wang:2022nqs,Karliner:2022erb, Meng:2022wgl, Zhang:2023teh,Yang:2023dzb}, and machine learning tools \cite{Ferretti:2021zis}. According to these studies,  $P_{\psi s}^{\Lambda}(4338)$  state can be identified as a $\Xi_c \bar{D}^{(*)}$ molecular structure whereas the $P_{\psi s}^{\Lambda}(4459)$ seems better to be a $\Xi^{(',*)}_c \bar{D}^{(*)}$ hadronic molecule. Apart from these models, different approaches arouse to elucidate inner structure of these states such as mixed configurations \cite{Chen:2020kco, Chen:2022wkh}, compact structure analysis~\cite{Li:2023aui, Maiani:2023nwj} and triangle singularities~\cite{Burns:2022uha}. The production and decay properties of these pentaquark states are also studied in Refs.~\cite{Liu:2020ajv, Cheng:2021gca, Yang:2021pio, Chen:2021tip, Wu:2021caw, Lu:2021irg,Paryev:2023icm}.

The aforementioned studies present information about the underlying structures of the $P_{\psi s}^{\Lambda}(4338)$ and $P_{\psi s}^{\Lambda}(4459)$ states. However, some models may not distinguish the color and spin configurations in the same multiplet. In order to shed light on the inner structure of these states, apart from radiative decays and weak decays, electromagnetic properties are needed to be obtained.  Studying electromagnetic properties of the related hadron paves the way for this task.  They encode valuable information about charge distribution and response to magnetic field.  Electromagnetic properties of $P_{\psi s}^{\Lambda}(4338)$ and $P_{\psi s}^{\Lambda}(4459)$ states are studied in Refs. 
\cite{Li:2021ryu, Gao:2021hmv, Ozdem:2021ugy, Wang:2022tib, Ozdem:2022kei, Ozdem:2022iqk,Zhang:2023teh, Ozdem:2023htj, Li:2024jlq, Ozdem:2024rqx, Li:2024wxr}. The magnetic moments of different exotic states are successfully obtained in Refs. \cite{Agamaliev:2016wtt,Ozdem:2017jqh,Ozdem:2017exj,Wang:2017dce,Ozdem:2023rkx}. For a good introduction of molecular interpretation about hadronic molecules, see Ref. \cite{Liu:2024uxn}.

Historically, the observation of hadrons containing charm valence quarks is followed by the identification of similar hadrons containing a bottom quark content. Accordingly, it  is natural to expect possible observation of the bottom analogues of the observed hidden-charm strange pentaquark states. Spectroscopic and electromagnetic properties together with strong and weak decays maintain valuable information for future experimental and  lattice QCD studies. Furthermore, theoretical and phenomenological studies may be helpful to get insights into the underlying structure and dynamics of these particles. This motivation led us to study electromagnetic properties of bottom analogues of hidden-charm strange pentaquark states. 

The paper is organized as follows: in Section \ref{wavefunc}, we introduce wave functions in molecular picture.  We briefly introduce magnetic moment concept in Section \ref{expression} and calculate magnetic moments in Section \ref{results}. Last section contains a brief discussion and concluding remarks.

\section{Wave Functions}\label{wavefunc} 
There are different pictures while studying hidden-heavy pentaquark states: molecular, diquark-diquark-antiquark, and diquark-triquark pictures. Among these pictures, molecular models are interesting since as outlined in Ref. \cite{Guo:2017jvc}, a clear-cut method for identifying a multiquark state (including hadronic molecule configurations) is to observe resonances decaying into a heavy quarkonium plus a light-quark baryon or meson with nonzero isospin. Studying electromagnetic properties of pentaquark states in molecular models presents useful clues about the internal structure. 

Wave function of a hadron can be written in terms of four degrees of freedom as
\begin{equation}
\Psi_{total}=\psi_{flavor}\psi_{spin}\psi_{color}\psi_{spatial},
\end{equation}
where the total wave function should be antisymmetric due to the Fermi statistics. We focus on $S$- wave hidden-bottom pentaquark states with strangeness $\mathcal{S}=-1$. In the ground state, color wave function $\psi_{color}$ is antisymmetric whereas  space wave function $\psi_{spatial}$ is symmetric. Therefore the symmetry requirement of spin-flavor is adequate for studying magnetic moments.

A pentaquark state in molecular picture is made up of a baryon and a meson. The possible quark arrangements are $(b\bar{b})(q_1 q_2 q_3)$ and $(\bar{b}q_3)(b q_1 q_2)$. In the former case, it is hard to form a loosely bound molecular state between bottomonium and a light baryon. Moreover, as mentioned in Ref. \cite{LHCb:2019kea}, narrow $P_c^+$ states could arise by binding a narrow baryon and a narrow meson such that separation of $c$ and $\bar{c}$ into distinct confinement volumes maintains a natural suppresion mechanism for the widths of $P_c^+$ states. Since the separation of $b$ and $\bar{b}$ is smaller then the separation of $c$ and $\bar{c}$ in charmonium, the same suppression may take place in the hidden-bottom pentaquark states. As a result, we take into account $(\bar{b}q_3)(b q_1 q_2)$ configuration and do not consider $(b\bar{b})(q_1 q_2 q_3)$ configuration in the molecular picture. 

According to aforesaid discussion, baryon inside the hidden-bottom strange pentaquark should be a singly-bottom baryon and meson inside the baryon should be bottom-light meson. The two light quarks inside the singly-bottom baryon can be either symmetric or antisymmetric. In the symmetric case,  $3_f$ flavor representation of bottom-light meson couples with the $6_f$ representation of the singly-bottom baryon to form $6_f \otimes 3_f = 10_f \oplus 8_{1f}$. In the antisymmetric case, the singly-bottom baryon has $\bar{3}_f$ flavor representation which couples to $3_f$ flavor representation of bottom-light meson to form $\bar{3}_f \otimes 3_f = 1_f \oplus 8_{2f}$. The so far observed pentaquark states are likely to be a member of octet states. For example, in the molecular scheme, the masses of $P_c(4380)$ and $P_c(4450)$ states with isospin $I=\frac{1}{2}$ are very close the mass threshold of $\bar{D}^{(*)} \Sigma_c^{(*)}$ and belong to the $8_{1f}$ representation since they were detected in the $J/\psi p$ channel \cite{Wang:2016dzu}. Therefore it is useful to investigate spectroscopic and electromagnetic properties of the hidden-heavy strange pentaquark states in molecular model \cite{Zou:2021sha}.

In Table \ref{tab:flavor}, we respectively list the flavor wave functions of  hidden-bottom strange molecular pentaquarks under $SU (3)$ symmetry.

\begin{table}[H]
\caption{\label{tab:flavor}The flavor wave functions of hidden-bottom strange molecular pentaquark states. $I$ represents isospin and $I_3$ represents its  third component. Here, $\{q_1q_2\}=\frac{1}{\sqrt 2}(q_1q_2+q_2q_1)$ and $[q_1q_2]=\frac{1}{\sqrt 2}(q_1q_2-q_2q_1)$.}
\begin{ruledtabular}
\begin{tabular}{cccc}
State&Flavor& $(I,I_3)$ & Wave function\\
\hline\\
$P_{bs}^1$& $8_{1f}$ & $(1,1)$ &$\sqrt{\frac{1}{3}}({\bar
b}u)(b\{us\})-\sqrt{\frac{2}{3}}({\bar b}s)(b\{uu\})$ \\  
& $8_{2f}$ &  &$({\bar b}u)(b[us])$ \\ \\ \hline \\

$P_{bs}^2$& $8_{1f}$ & $(1,0)$ &$\sqrt{\frac{1}{6}}[({\bar b}d)(b\{us\})+({\bar b}u)(b\{ds\})]-\sqrt{\frac{2}{3}}({\bar b}s)(b\{ud\})$ \\  

&  $8_{2f}$&  &$\frac{1}{\sqrt2}\{ ({\bar b}d)(b[us])+({\bar b}u)(b[ds]) \}$  \\ \\ \hline \\

$P_{bs}^3$& $8_{1f}$ & $(1,-1)$ &$\sqrt{\frac{1}{3}}({\bar
b}d)(b\{ds\})-\sqrt{\frac{2}{3}}({\bar b}s)(b\{dd\})$ \\  

& $8_{2f}$ &  &$({\bar b}d)(b[ds])$ \\ \\  \hline \\

$P_{bs}^4$& $8_{1f}$ & $(0,0)$ &$\sqrt{\frac{1}{2}}[({\bar b}u)(b\{ds\})-({\bar b}d)(b\{us\})]$ \\  

& $8_{2f}$ &  &$\frac{1}{\sqrt6} \{ ({\bar b}d)(b[us])-({\bar b}u)(b[ds])-2(\bar b s)(b[ud]) \}$ \\ \\  \hline \\

$P_{bss}^1$& $8_{1f}$ & $(\frac{1}{2},\frac{1}{2})$ &$\sqrt{\frac{1}{3}}({\bar b}s)(b\{us\})-\sqrt{\frac{2}{3}}({\bar b}u)(b \{ss\})$ \\  

& $8_{2f}$ &  &$({\bar b}s)(b[us])$ \\ \\  \hline \\

$P_{bss}^2$& $8_{1f}$ & $(\frac{1}{2},-\frac{1}{2})$ &$\sqrt{\frac{1}{3}}({\bar b}s)(b\{ds\})-\sqrt{\frac{2}{3}}({\bar b}d)(b\{ss\})$ \\  

& $8_{2f}$ &  &$({\bar b}s)(b[ds])$ \\  

\end{tabular}
\end{ruledtabular}
\end{table}

One can understand these flavor representations as follows:  $8_{1f}$ flavor representations are  combinations or mixings of two states whereas $8_{2f}$ flavor representations are pure states except $I_3=0$.

\section{Magnetic Moments}\label{expression}
Since quarks are Dirac fermions, at the quark level the operators of the magnetic moments are
\begin{equation}
\hat{\mu}= \sum_i \frac{Q_i}{2m_i}\hat{\sigma}_{i},
\end{equation}
where $Q_i$, $m_i $, and $\hat{\sigma}_{i}$ denote charge, mass, and Pauli’s spin matrix of the $i$th quark, respectively. The total magnetic moment formula for the $S$-wave molecular pentaquarks can be written as
\begin{eqnarray}
\hat{\mu}  = \hat{\mu}_{B}+\hat{\mu}_{M},
\end{eqnarray}
where the  subscripts $B$ and $M$ represent the baryon and meson, respectively. The baryon and meson magnetic moments are
\begin{eqnarray}
		\mu  &=& \langle \psi  | \hat{\mu}_{B}+\hat{\mu}_{M}| \psi  \rangle \nonumber\\
		&=&
		\sum_{SS_z,ll_z} \langle  SS_z,ll_z|JJ_z \rangle^{2}  \left \{ 
		\sum_{S_B,S_M}\ \langle\ S_B \widetilde{S}_{B},S_M \widetilde{S}_{M}|SS_z\ \rangle^{2} \Bigg [
		\widetilde{S}_{M}\bigg(\mu_{\bar{b}} + \mu_{q_3}\bigg )\nonumber\right.\\
		&+&\left.
		\sum_{\widetilde{S}_{b}}\ \langle\ S_b \widetilde{S}_{b},S_{Di} \widetilde{S}_{B}-\widetilde{S}_{b}|S_B \widetilde{S}_{B}\rangle^{2}\bigg(g\mu_{b}\widetilde{S}_{b}+(\widetilde{S}_{B}-\widetilde{S}_{b})(\mu_{q_{1}}+\mu_{q_{2}})\bigg )
		\Bigg ]\right \},          
\end{eqnarray}
where $\psi$ represents the flavor wave function in molecular model listed in in Table \ref{tab:flavor}, $S_M$, $S_B$, $S_{Di}$, $S_b$ are the meson, baryon, the diquark spin inside the baryon, and bottom quark, respectively. $\widetilde{S}$ is the third spin component.

We take $8_{2f}$ of $P_{bs}^4$ hidden-bottom strange molecular state for an example. Let us take $J^P=\frac{1}{2} (\frac{1}{2}^+ \otimes 1^-)$ where $(J_B^{P_b} \otimes J_M^{P_m})$ correspond to the angular momentum and parity of baryon and meson. The wave function
\begin{equation}
\vert P_{bs}^4 \rangle = \frac{1}{\sqrt6} \{ ({\bar b}d)(b[us])-({\bar b}u)(b[ds])-2(\bar b s)(b[ud]) \},
\end{equation}
gives magnetic moment as
\begin{align}
\mu  & =
\langle P_{bs}^4 | \hat{\mu}_{B}+\hat{\mu}_{M}\ | P_{bs}^4  \rangle\nonumber\\
& =\langle \frac{1}{2}\frac{1}{2},1 0 |\frac{1}{2}\frac{1}{2}\rangle^{2}\Bigg [
\langle \frac{1}{2}\frac{1}{2},0 0 |\frac{1}{2}\frac{1}{2}\rangle^{2}
\Bigg 	(\frac{1}{6}(\frac{1}{2}g\mu_{b})+\frac{1}{6}(\frac{1}{2}g\mu_{b})+\frac{4}{6}(\frac{1}{2}g\mu_{b})\Bigg )\Bigg ]
+\nonumber\\ & 
\ \ \ \ \ \langle \frac{1}{2}-\frac{1}{2},1 1 |\frac{1}{2}\frac{1}{2}\rangle^{2}\Bigg [	\Bigg 	(\frac{1}{6}(\frac{1}{2}g\mu_{\bar{b}}+\frac{1}{2}g\mu_{d})+\frac{1}{6}(\frac{1}{2}g\mu_{\bar{b}}+\frac{1}{2}g\mu_{u})+\frac{4}{6}(\frac{1}{2}g\mu_{\bar{b}}+\frac{1}{2}g\mu_{s})\Bigg )
+\nonumber\\ & 
\ \ \ \ \ 	\langle \frac{1}{2}-\frac{1}{2},0 0 |\frac{1}{2}-\frac{1}{2}\rangle^{2} 	\Bigg 	(\frac{1}{6}(-\frac{1}{2}g\mu_{b})+\frac{1}{6}(-\frac{1}{2}g\mu_{b})+\frac{4}{6}(-\frac{1}{2}g\mu_{b})\Bigg )\Bigg ]
\nonumber\\ 
& = \frac{1}{9} (\mu_{u}+\mu_{d}+4\mu_{s}-9\mu_{b} ).
\end{align}

We use the following constituent quark masses for numerical analysis \cite{Thakkar:2016dna}
\begin{eqnarray}
m_u \ = 0.338 \ \mbox{GeV}, \ m_d \ =\  0.350\ \mbox{GeV}, \ 	m_s \ =\ 0.500\ \mbox{MeV},\  \ m_b \ =\ 4.67\ \mbox{GeV}.
\end{eqnarray}

\section{Numerical Results and Discussion}\label{results}
Before calculating magnetic moments of hidden-bottom strange pentaquarks, it will be beneficial to calculate magnetic moments of octet baryons which are experimentally observed. In Table \ref{tab:magmomexp}, we present our results with the ones observed in experiments. Our results are in rough agreement with the experimental results. 

\begin{table}[H]
\caption{ \label{tab:magmomexp}Magnetic moments of the octet baryons. The results are listed in unit of nuclear magneton $\mu_N$.}
\begin{ruledtabular}
\begin{tabular}{cccc}
State&Magnetic moment& This work & Experiment \cite{Workman:2022ynf} \\
\hline \\
$p$&  $\frac{4}{3}\mu_u-\frac{1}{3}\mu_d$ & $2.764$  &$2.793$  \\ \\ 
$n$&  $\frac{4}{3}\mu_d-\frac{1}{3}\mu_u$ & $-1.808$  &$-1.913$  \\ \\
$\Lambda$&  $\mu_s$  & $-0.625$  &$-0.613 \pm 0.006$  \\ \\
$\Sigma^+$&  $\frac{4}{3}\mu_u-\frac{1}{3}\mu_s$   & $2.675$  &$2.460 \pm 0.006$ \\ \\
$\Sigma^-$ & $\frac{4}{3}\mu_d-\frac{1}{3}\mu_s$ & $-0.982$  &$-1.160 \pm 0.025$  \\ \\
$\Xi^0$  & $\frac{4}{3}\mu_s-\frac{1}{3}\mu_u$  & $-1.450$  &$-1.250 \pm 0.014$  \\ \\
$\Xi^-$  &   $\frac{4}{3}\mu_s-\frac{1}{3}\mu_d$ & $-0.536$ &$-0.651\pm 0.0025$ \\ \\ 
$\Omega^-$   &  $3\mu_s$ & $ -1.876$ &$-2.020\pm 0.05$  \\ 

\end{tabular}
\end{ruledtabular}
\end{table}

Motivated with this picture, we calculate magnetic moments of hidden-bottom strange pentaquarks. The results are presented in Table \ref{tab:magmomhbs}.

\begin{table}[H]
\caption{ \label{tab:magmomhbs}Magnetic moments of the molecular hidden-bottom strange pentaquark states. The results are listed in unit of nuclear magneton $\mu_N$.}
\begin{ruledtabular}
\begin{tabular}{ccccc}
State& $J_B^{P_b} \otimes J_M^{P_m}$ & $I(J^P)$ & Magnetic moment $(8_{1f})$ &  Magnetic moment $(8_{2f})$ \\
\hline \\
$P_{bs}^1$& $\frac{1}{2}^+ \otimes 0^-$  &  $1(\frac{1}{2})^-$ &$1.939$ & $-0.067$\\ \\ 
& $\frac{1}{2}^+ \otimes 1^-$  &  $1(\frac{1}{2})^-$ &$-0.468$ & $1.300$\\ \\
&   &  $1(\frac{3}{2})^-$ &$2.206$ & $1.850$\\ \\ \hline \\

$P_{bs}^2$& $\frac{1}{2}^+ \otimes 0^-$  &  $1(\frac{1}{2})^-$ &$0.415$ & $-0.067$\\ \\ 
& $\frac{1}{2}^+ \otimes 1^-$  &  $1(\frac{1}{2})^-$ &$-0.265$ & $0.386$\\ \\
&   &  $1(\frac{3}{2})^-$ &$0.224$ & $0.478$\\ \\ \hline \\

$P_{bs}^3$& $\frac{1}{2}^+ \otimes 0^-$  &  $1(\frac{1}{2})^-$ &$-1.109$ & $-0.067$\\ \\ 
& $\frac{1}{2}^+ \otimes 1^-$  &  $1(\frac{1}{2})^-$ &$-0.062$ & $-0.573$\\ \\
&   &  $1(\frac{3}{2})^-$ &$-1.757$ & $-0.893$\\ \\ \hline \\

$P_{bs}^4$& $\frac{1}{2}^+ \otimes 0^-$  &  $0(\frac{1}{2})^-$ &$-0.076$ & $-0.067$\\ \\ 
& $\frac{1}{2}^+ \otimes 1^-$  &  $0(\frac{1}{2})^-$ &$0.389$ & $-0.105$\\ \\
&   &  $0(\frac{3}{2})^-$ &$0.470$ & $-0.257$\\ \\ \hline \\

$P_{bss}^1$& $\frac{1}{2}^+ \otimes 0^-$  &  $\frac{1}{2}(\frac{1}{2})^-$ &$-0.261$ & $-0.067$\\ \\ 
& $\frac{1}{2}^+ \otimes 1^-$  &  $\frac{1}{2}(\frac{1}{2})^-$ &$0.815$ & $-0.350$\\ \\
&   &  $\frac{1}{2}(\frac{3}{2})^-$ &$0.741$ & $-0.625$\\ \\ \hline \\

$P_{bss}^2$& $\frac{1}{2}^+ \otimes 0^-$  &  $\frac{1}{2}(\frac{1}{2})^-$ &$-0.871$ & $-0.067$\\ \\ 
& $\frac{1}{2}^+ \otimes 1^-$  &  $\frac{1}{2}(\frac{1}{2})^-$ &$-0.201$ & $-0.350$\\ \\
&   &  $\frac{1}{2}(\frac{3}{2})^-$ &$-1.608$ & $-0.625$\\  

\end{tabular}
\end{ruledtabular} 
\end{table}

As can be seen in the table, the magnetic moments of the $8_{1f}$ and $8_{2f}$ flavor representations are different: no magnetic moment of $8_{1f}$ flavor representation is equal to magnetic moment of $8_{2f}$ flavor representation. We highlight our results as follows:

\begin{itemize}

\item In the $P_{bs}^1$ states, the possible observation of magnetic moment may serve to estimate the quantum numbers. According to the our results, observation of a negative magnetic moment may support $J^P=1(\frac{1}{2})^-$ assignment. Furthermore, a magnetic moment measurement may also serve for determining the angular momentum of baryon and meson in the molecular pentaquark state, $\frac{1}{2}^+ \otimes 0^-$ or  $\frac{1}{2}^+ \otimes 1^-$. The observation of the magnetic moment around $-0.5 ~ \mu_N$ may give a hint about the internal quantum numbers. $J^P=\frac{3}{2}^-$ have the largest magnetic moments in both $8_{1f}$ and  $8_{2f}$ flavor representations. 

\item In the $P_{bs}^2$ states, the sign of the magnetic moment for the quantum number $J^P=\frac{1}{2}^-$ has opposite signs in $8_{1f}$ and  $8_{2f}$ flavor representations. A possible observation of magnetic moment may identify the angular momentum configuration. $J^P=\frac{3}{2}^-$ of  $8_{2f}$  flavor representation has the largest magnetic moment in magnitude.

\item All the magnetic moments of $P_{bs}^3$ states have negative sign. Therefore measurement of magnetic moment may identify the quantum number and spin configurations of these states.  $J^P=\frac{3}{2}^-$ of  $8_{1f}$  flavor representation has the largest magnetic moment.

\item In the $P_{bs}^4$ states, the quantum numbers $J^P=\frac{1}{2}^-$ and $J^P=\frac{3}{2}^-$ of $\frac{1}{2}^+ \otimes 1^-$ spin configuration in $8_{1f}$ flavor representation has positive magnetic moments whereas the rest has negative magnetic moments. A possible observation of positive value magnetic moment may identify spin configuration.  $J^P=\frac{3}{2}^-$ of  $8_{1f}$  flavor representation has the largest magnetic moment in magnitude.

\item The  quantum number $J^P=\frac{1}{2}^-$ of  $\frac{1}{2}^+ \otimes 0^-$  spin configuration has negative magnetic moments in  $8_{1f}$ and  $8_{2f}$ flavor representations of  $P_{bss}^1$ states. $J^P=\frac{1}{2}^-$ of  $\frac{1}{2}^+ \otimes 1^-$  spin configuration has positive magnetic moment in $8_{1f}$ flavor representation whereas the magnetic moment is negative in the $8_{2f}$ flavor representation. This is also valid for the quantum number $J^P=\frac{3}{2}^-$. Unlike  previous cases, $J^P=\frac{1}{2}^-$ of  $\frac{1}{2}^+ \otimes 1^-$ in $8_{1f}$ flavor representation has the largest magnetic moment in magnitude.

\item In the $P_{bss}^2$ states, all the magnetic moments have negative sign. Therefore measurement of magnetic moment may identify the quantum number and spin configurations of these states. $J^P=\frac{3}{2}^-$ of  $8_{1f}$  flavor representation has the largest magnetic moment.

\item One of the interesting patterns is that all the $J^P=\frac{1}{2}^-$ of $\frac{1}{2}^+ \otimes 0^-$ spin configurations in the $8_{2f}$  flavor representation has the same magnetic moment, $\mu=-0.067 \mu_N$. The reason for this could be that magnetic moments of these states are only determined by $b$ quark. 

\item The other interesting pattern is that the magnetic moments of $8_{2f}$  flavor representations of $P_{bss}^1$ and $P_{bss}^2$ states are same for each corresponding quantum numbers and spin configurations. Apart from  $J^P=\frac{1}{2}^-$ of $\frac{1}{2}^+ \otimes 0^-$ spin configuration in $8_{2f}$ of all states, magnetic moments could be determined by $s$ and $b$ quarks in the $\frac{1}{2}^+ \otimes 1^-$ spin configurations.

\end{itemize}

Detailed analysis showed that the contribution of light quarks to the magnetic moment cancel each other in some configurations. The hadron contains not only valence quarks but also sea quarks, whose effects are embodied in the constituent quark masses. Contributions of sea quarks to the magnetic moment are studied in unquenched quark model \cite{Bijker:2009up,Santopinto:2010zza,Bijker:2012zza}. These results showed that the inclusion of the effects of sea quarks preserves the results of constituent quark model for the magnetic moments of octet baryons.

\section{Final Remarks}\label{final}
The unabated pursuit for the discovery of new exotic hadron states may provide some challenges for the inner structures. Although more than 20 years passed over the discovery of $\chi_{c1}(3872)$, there is still no consensus about the inner structure of this state. 

The mass of a hadron is an important quantity since one can compare it to corresponding thresholds. This may give an evidence for the stability of that state. The magnetic moment of a hadron is an important quantity as its mass. It contains valuable information about the inner structures. Motivated by the observation of hidden-charm strange pentaquarks, in this work we investigate the magnetic moments of hidden-bottom strange pentaquark states in molecular picture. 

Our results reflect some interesting features. For example, in all hidden-bottom strange pentaquark states, $J^P=\frac{3}{2}^-$ of $\frac{1}{2}^+ \otimes 1^-$ has the largest magnetic moments, except in the $P_{bss}^1$ states where the $J^P=\frac{1}{2}^-$ of $\frac{1}{2}^+ \otimes 1^-$ has the largest moment. The observation of hidden-bottom strange pentaquarks with strangeness-2 may have some other interesting properties. Our calculations may help to determine the corresponding quantum numbers and spin configurations of hidden-bottom strange pentaquark states.

In addition to magnetic moments, decay modes, decay channels and branching ratios are also important for determining the inner structure.  The magnetic moments calculated here could be compared with other studies. We encourage three-point QCD sum rule, light-cone QCD sum rule or lattice QCD studies for comparison, since using different models for the same task may help for understanding and elucidating the problem in hand. Lattice QCD studies consume highly large times and therefore need advanced computational resources such as supercomputing centres. The magnetic moments can also be calculated by the QCD sum rules formalism where the bare quark masses are chosen as inputs for quark masses. In constituent quark models, the quark masses are different than bare quark masses. Quark masses are not accesible to direct experimental measurements. From a physical perspective, the calculated quantities (magnetic moments in this task) are identical, thus indicating that the outcomes should be comparable. However, given that the two methodologies employ distinct approximations and utilize distinct parameter sets, it is probable that disparate outcomes will be observed. Only experimental outcomes may serve to elucidate which model offers more consistent and superior outcomes, thereby enabling the differentiation between these models. 

We hope that our results may be helpful for theoretical and experimental studies on the hidden-bottom strange pentaquark states. We also hope that this research would contribute our understanding of the molecular states in hadron physics.

\bibliography{hbs-pentaquark-r1}

\end{document}